\documentstyle[psfig,epsf]{article}
\begin{document}

\begin{center}     
{\Large \bf On the numerical technique of Casimir energy calculation} \\

\vspace{4mm}

Il.  Malakhov$^{a}$, P. Silaev$^{a,b}$, K.  Sveshnikov$^{a,b}$\\
$^a$Physics Department and $^{b}$Institute of Theoretical Micro Physics\\
Moscow State University, Moscow 119992, Russia\\ 

\end{center}

\begin{abstract}
A non-subtractive recipe of Casimir energy renormalization efficient in 
the presence of logarithmically divergent terms is proposed.  It is 
demonstrated that it can be applied even in such cases, when energy levels 
can be obtained only numerically whereas neither their asymptotical 
behavior, nor the analytical form of the corresponding spectral equation 
can be studied.  The results of numerical calculations performed with this 
method are compared to those obtained by means of explicit subtraction of 
divergent terms from energy.

\end{abstract}

\section{Introduction}
   
Ever since Casimir \cite{casimir} has obtained corrections to the energy of a
macroscopic system due to vacuum fluctuations of quantized electromagnetic 
field in 1948 this effect has been intensively studied both from 
theoretical and experimental points of view. Nevertheless the calculation 
of Casimir energy except for the most simple problems involving free 
fields inside cavities with flat boundaries remains quite non-trivial yet. 
To realize this, one can recall a great number of papers devoted to a free 
field confined in the interior of a sphere \cite{boyer}--\cite{blau}. 
Despite the fact that in this case spectral equations can be written out 
explicitly, the first analytical results for massive scalar field have 
been obtained only in \cite{bordag_1}. The analogous problem for fermions 
has been solved in \cite{elizalde}.

It should be stressed that the knowledge of analytical form of spectral 
equation has been crucial for the employment of the method proposed in 
\cite{bordag_1} since it makes possible the transition from the sums 
containing the unknown energy levels to the integrals with the explicit 
integrands \cite{bordag_2}. The main goal of this paper is to propose a 
method which can be applied to numerical calculation of Casimir energy in 
situations when these requirements are not met. Another goal is to 
demonstrate that it's possible to perform all the necessary calculations 
without employment of a rather standard trick \cite{vepstas}, 
\cite{bordag_3}, which lets one overcome problems arising due to the 
presence of logarithmic divergency in Casimir energy of free massless 
fields inside spherical shells. Note that logarithmic divergency appears 
as a consequence of a curved surface bounding the shell and makes the 
energy renormalization ambiguous. The main idea of the mentioned trick is 
to consider the ``inner'' and ``extra'' problems together since their 
logarithmic divergencies cancel each other.

The ambiguity of Casimir energy renormalization in the presence of 
logarithmic divergency is quite obvious. Indeed, in case of massless 
fields the energy of the system can be characterized by a single 
dimensional parameter $L$ which is the linear size of the system. The 
regularization parameter $\alpha$ can be also chosen to have a dimension 
of length. In the absence of logarithmic divergency the ``minimal 
subtraction'' of singular terms is not only natural but also well 
grounded.  Indeed, any divergent term in the expansion of regularized 
energy, which is proportional to $\alpha^{-s}$ ($s>0$) is inevitably 
proportional to $L^{s-1}$, i.e. to the non-negative power of $L$. This 
makes it possible to normalize the final result at $L=\infty$, where 
Casimir energy should become zero, and subtract all singular terms at the 
same time.  After such subtraction the only remaining term in the limit 
$\alpha\to 0$, which reads $c\alpha^0/L$, provides the final result.

In the presence of logarithmic divergency the subtraction becomes 
ambiguous, since in order to renormalize the term $c \alpha^0 
\log(\alpha/L)/L$, one should subtract $c \alpha^0 \log(d \alpha 
/L)/L$, where $d$ is an arbitrary constant, which cannot be determined 
from the normalizing condition at $L\to \infty$. 

\section{Massive scalar field in 1D}

To illustrate the main idea of technique under investigation let's study 
Casimir energy with the logarithmic divergency in the most trivial case, 
i.e. Casimir energy of massive scalar field on an interval of length $L$ 
with Dirichlet boundary conditions at the ends of the interval. It reads:  
$$ {\cal E}_{cas}={1\over 2}\sum_{n=1}^\infty  \omega_n ={1\over 2} 
\sum_{n=1}^\infty  \sqrt{(\pi n/L)^2+m^2} \eqno (1)   $$

\noindent First of all, the renormalization of Casimir energy is performed 
without intermediate subtraction of Minkowski vacuum contribution 
prescribed by ``standard'' approaches.  The regularization of (1) 
requires the introduction of parameter $\alpha$ having a dimension of 
length, which stands in the argument of the cut-off function $F(\alpha 
\omega_n)$:

$$ {\cal E}_{cas}^{(r)}={1\over 2}\sum_{n=1}^\infty  \omega_n F (\alpha 
\omega_n) \eqno (2)$$

\noindent Trivial considerations based on dimensional analysis lead to the 
following expression for the regularized Casimir energy 
$${\cal E}_{cas}^{(r)}\simeq c_{-2}{ L \over \alpha^2} + c_{-1}{ L^0 \over 
\alpha^1} + c_{0}{ 1 \over L} + c_{\lambda}{ m^2 L \log(\alpha/L) } + 
\cdots \eqno (3)$$

\noindent It can be easily verified that for various cut-off functions 
such as $F(x)=\exp(-x)$, $F(x)=\exp(-x^2)$, $F(x)=\exp(-x^3)$, $\ldots$, 
$F(x)=\exp(-x^6)$, $F(x)=\exp(-2\cosh(x)+2)$, $\ldots$, identical 
$c_{\lambda}$ are obtained, while $c_0$ are different. The identity of 
$c_{\lambda}$ for different $F(x)$ can be demonstrated by means of the 
following estimation for the sum giving rise to the logarithmic 
divergency:  

$$ {1\over 2}\sum_{n=1}^\infty {1\over 2} {m^2\over \omega_n 
} F (\alpha \omega_n) \sim  {m^2 L\over 4\pi } \log N \sim {m^2 L\over 
4\pi } \log (\Delta x L/\alpha) =$$ $${m^2 L\over 4\pi } \log (L/\alpha)  
+ {m^2 L\over 4\pi } \log (\Delta x) \; ,  \eqno (4)$$ 

\noindent where $N\sim \Delta x L/\alpha$ and $\Delta x$ is a cut-off 
interval of $F(x)$.

Since any subtraction in the presence of logarithmic divergency is 
ambiguous this procedure should be excluded from consideration along with 
the logarithmic divergency itself. To achieve that one should calculate 
$\partial_L^2 {\cal E}_{cas}^{(r)}$:

$$\partial_L^2 {\cal E}_{cas}^{(r)}\simeq c_{0}{ 2 \over L^3} + 
c_{\lambda}{ m^2 \over L } + \cdots \eqno (5) $$

\noindent The obtained expression is regular in the limit $\alpha \to 0$, 
so no subtraction is required. The knowledge of the function $\partial_L^2 
{\cal E}_{cas}^{(r)}$, lets one reconstruct the required ${\cal 
E}_{cas}^{(r)}$ unambiguously, since the initial conditions at 
$L\to\infty$ are well-known:  both ${\cal E}_{cas}^{(r)}$ and $\partial_L 
{\cal E}_{cas}^{(r)}$ should become zero in this limit.  Note that while 
(5) doesn't describe the asymptotical behavior of $\partial_L^2 {\cal 
E}_{cas}^{(r)}$ at $L\to\infty$, it demonstrates the disappearance of all 
singular terms in this quantity. Moreover the following integral 
approximation shows that $\partial_L^2 {\cal E}_{cas}^{(r)}$ does really 
vanish for $L \to \infty$: 

$$ {\cal E}_{cas}^{(r)}={1\over 2}\sum_{n=1}^\infty 
\omega_n F (\alpha \omega_n) = {1\over 4}\sum_{n=-\infty}^\infty  \omega_n 
F (\alpha \omega_n) -{m\over 4} F(\alpha m) \approx $$  $$ \approx {1\over 
2} \int_{-\infty}^\infty dx (L/\pi) \sqrt{x^2+m^2} F (\alpha 
\sqrt{x^2+m^2}) -{m\over 4} F(\alpha m) \eqno(6) $$

\section{Method of calculation in general case}

The proposed method can be generalized in such a way that it doesn't 
require the analytical expression for energy levels for its application.  
Suppose one has a set of energy levels  of some spectrum $\omega_n$ and 
the corresponding regularized Casimir energy contains the logarithmically 
divergent term.  First of all it turns out to be possible to modify the 
initial expression for the Casimir energy by introduction of some 
parameter $\mu$ in such a way that 

$$ {1\over 2}\sum_{n=1}^\infty  \sqrt{\omega_n^2-\mu^2} F (\alpha 
\sqrt{\omega_n^2-\mu^2}) \eqno (7)$$

\noindent no longer contains the logarithmic divergency. For the massive 
scalar field on an interval $\mu$ obviously equals to the mass of the 
field. In less trivial three-dimensional cases with spherical symmetry 
$\mu$ is some parameter having a dimension of mass which characterizes the 
total coefficient by the logarithmic divergency with all values of angular 
momentum taken into account.

The next step is to introduce an ``additional mass'' of the field ${\cal 
M}$ and study the behavior of Casimir energy in the range from ${\cal 
M}=0$ to ${\cal M}=\infty$. In fact it's convenient to deal with another 
parameter $M$, which is related to ${\cal M}$ by means of  $M^2 \equiv 
{\cal M}^2+\mu^2$, and study the ``modified'' Casimir energy

$$ {\cal E}_{cas} (M) = {1\over 2}\sum_{n=1}^\infty \sqrt{\omega_n^2+ 
M^2-\mu^2} \eqno(8)$$

\noindent as a function of $M$ in the range from $M=\infty$ to $M=\mu$. To 
carry this out, one should calculate numerically the following quantity in 
the specified range of $M$:

$$ \partial^2_M  \left( {\cal E}_{cas}^{(r)} ({M})/M \right) =$$

$$ = \partial^2_M  \left[ {1\over 2}\sum_{n=1}^\infty  \sqrt{ 
{\omega_n^2-\mu^2 \over M^2} + 1\; } \;\; F \left(\alpha \sqrt{ 
{\omega_n^2-\mu^2 \over M^2} +1\;} \, \right) \right] \eqno(9) $$

\noindent Note that in contrast to our previous considerations the 
dimensionless quantity

$$ {1 \over M}  \sqrt{ \omega_n^2 + M^2-\mu^2   \; }= \sqrt{ 
{\omega_n^2-\mu^2 \over M^2} + 1\; } $$

\noindent has been substituted to the argument of $F(x)$, so that $\alpha$ 
should be also taken dimensionless.

An alternative interpretation  of (9) follows from the observation that 
$M$ acts just as an effective length $L$ in the expression for Casimir 
energy.  Indeed, (9) can be obtained from the initial expression as a 
result of the following sequence of transformations of the spectrum. At 
the first step $\omega_n$ is transformed to $\omega_n' = \sqrt{ 
\omega_n^2-\mu^2 }/\mu$, which is the dimensionless form of the spectrum 
with the subtracted effective mass.  After that the scale transformation 
of the system $x\to x (M/\mu)$ and $\omega_n'\to \omega_n'' =\omega_n' 
/(M/\mu)$ is performed.  In the end the unit mass is ``added'' to the 
obtained spectrum:

$$\omega_n''\to \sqrt{ (\omega_n')^2+1}= \sqrt{ {\omega_n^2-\mu^2 \over 
M^2} + 1\; } \eqno(10)$$

\noindent The limit $M\to\infty$ obviously corresponds to the infinite 
size of the system, while for $M=\mu$ one obtains the initial spectrum 
divided by $\mu$.
                                                        
It's easy to see that all divergent terms in (9) vanish. In the limit 
$n\to \infty$ one can make use of the following expansion $$ \sqrt{ 
{\omega_n^2-\mu^2 \over M^2} + 1\; }\approx { \sqrt{ \omega_n^2-\mu^2} 
\over M } + { M \over 2 \sqrt{ \omega_n^2-\mu^2} } + \cdots \eqno(11)$$ 

\noindent The first term in this expansion gives rise to the sum which is 
free of logarithmic divergency owing to the definition of $\mu$. Other 
divergencies are proportional to $(M/\alpha)^2/M$ and $(M/\alpha)^1/M$ 
and vanish when the second-order derivative is taken.  The second term 
leads to the logarithmic divergency with the coefficient by it 
proportional to $M^1$, which also vanishes while taking the second-order 
derivative.

As a result one has the expression (9) regular in the limit $\alpha\to 0$ 
and the natural normalizing condition ${\cal E}_{cas} ( M \to \infty)=0$.  
The latter can be understood from two different points of view. On one 
hand the quantized field with the infinitely large mass should definitely 
have zero Casimir energy. On the other hand the Casimir energy in the 
limit of the infinite size of the system should become zero. Whichever 
interpretation is chosen, these two points let one reconstruct the 
required ${\cal E}_{cas} (M=\mu)$ corresponding to the non-modified 
spectrum.

Note, that principally one could consider ${\cal E}_{cas}^{(r)} ({M})$ 
instead of ${\cal E}_{cas}^{(r)} ({M})/M$. However that would increase the 
order of derivative  required to exclude all divergent terms what is 
undesirable from the point of view of real-time computations.

The proposed method turns out to be efficient not only in the most trivial 
one-dimensional cases, but also in more realistic three-dimensional ones. 
However to employ it in three-dimensional case one should inevitably 
calculate the fourth-order derivative of the Casimir energy (9) since 
the main singular term, which is proportional to the volume of the system, 
reads $ c_{-4} L^3/\alpha^4 $. It should be also noted that in this case 
the summations become more lengthy and  sophisticated. Typically the final 
result turns out to be about 40 orders lower than the values of partial 
sums obtained in the process of its calculation. As a consequence, extra 
floating-point precision is required. 

\vskip 5 true mm

\section{Numerical results}

\vskip 2 true mm
  
For scalar field on an interval $[0;L]$ with $L=1$ the spectrum reads
$$\omega_n = \sqrt{ { (\pi n)^2 \over L^2 } + m^2 } \eqno (12)$$

\begin{figure}[htb]
\centerline{\psfig{figure=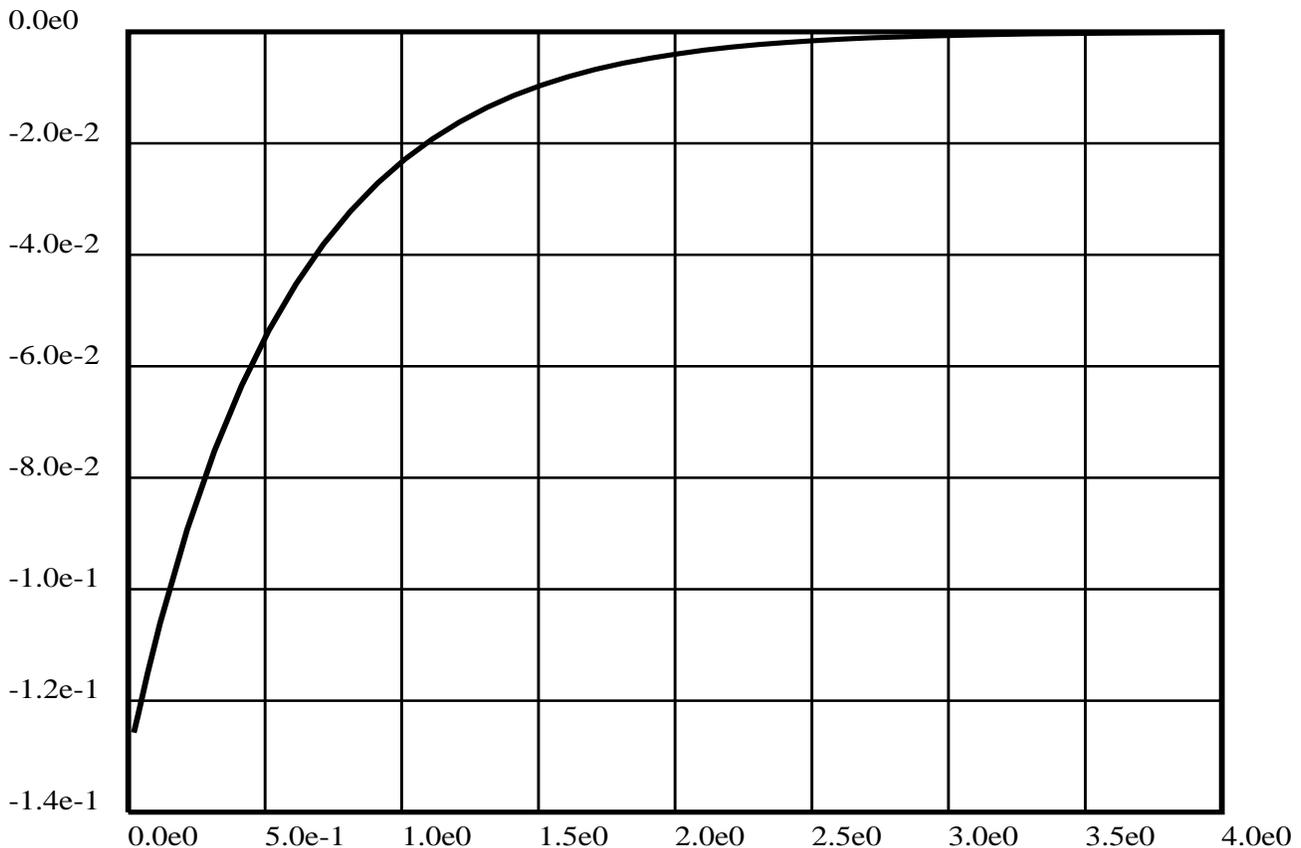,height=12cm,width=18cm}}
\caption{\label{cas_segm} The Casimir energy of the massive scalar field 
on a unit interval as a function of the mass of the field.}
\end{figure}

The result of straightforward application of the proposed technique with 
various cut-off functions such as $F(x)=\exp(-x)$, $F(x)=\exp(-x^2)$, 
$F(x)=\exp(-x^3)$, $\ldots$, $F(x)=\exp(-x^6)$, $F(x)=\exp(-2\cosh(x)+2)$ 
is presented on Fig.~\ref{cas_segm}. It has been shown that for each of 
these functions the same result is obtained and what's more the precision 
of coincidence depends only on the number of energy levels taken into 
account and the number of right digits used in the realization of 
floating-point arithmetics as well.

As to dependence of the Casimir energy on the mass of the field some 
important aspects should be stressed. First of all in the limit $m \to 0$ 
a well-known result for the massless scalar field is obtained.  In the 
range of large values of $m$ Casimir energy decreases exponentially as 
$e^{-2mL}$ what could be expected from qualitative considerations. To 
summarize, the results obtained in this trivial case are in full agreement 
with the previously known results, obtained with the traditional 
technique, based on an explicit subtraction of divergent terms.

To demonstrate how this approach can be employed in less trivial cases it 
makes sense to consider the massless scalar field inside of a spherical 
shell of radius $R=1$. In this case the same set of cut-off functions has 
been exploited. The values of an effective mass $\mu=0.1377$ obtained with 
each of these functions coincide up to the first four digits. Consequently 
the precision of the obtained $E_{cas}=3.790\cdot 10^{-3}$ has the same 
order, what corresponds to about 200 $s$-levels taken into account while 
performing the calculations. In fact, the number of energy levels taken 
into account is directly affected by the range which the regularization 
parameter $\alpha$ used in the calculations belongs to. Therefore one can 
control precision of the final result simply changing the range of 
employed values of the regularization parameter.

\begin{figure}[thb]
\centerline{\psfig{figure=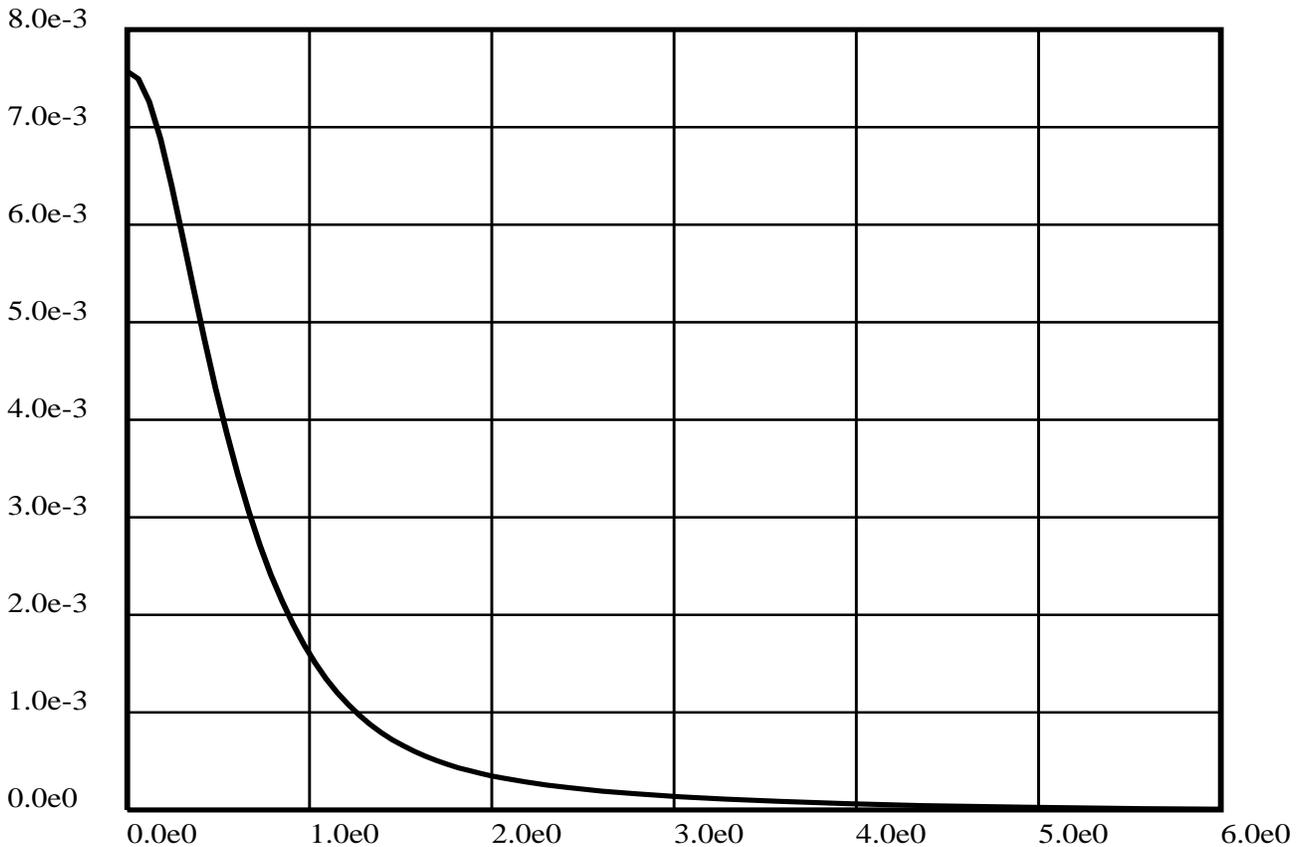,height=12cm,width=18cm}}
\caption{\label{sph_sc} The Casimir energy of the scalar field inside of a 
spherical shell of radius 1 obeying Dirichlet boundary conditions as a 
function of the mass of the field.}
\end{figure}

Note that in the framework of this approach not only Casimir energy of the 
massless scalar field (corresponding to ${\cal M}=0$) has been obtained 
but also Casimir energy for all possible mass values in the range from 
zero to the ``effective'' infinity. The dependence of the Casimir energy 
of the scalar field inside the sphere on the mass of the field is 
presented on Fig.~\ref{sph_sc}.

It should be stressed that on the contrary to the results obtained with 
the traditional subtractive technique in \cite{bordag_1} our result 
doesn't contain logarithmical singularity at ${\cal M}=0$ what seems more 
reasonable from the physical point of view. The most likely explanation of 
this discrepancy is that the subtractive procedure may very well contain 
some arbitrariness, even if one puts the proper normalizing condition. As 
a result some function which has regular behavior at ${\cal M} \to \infty$ 
(and thus doesn't violate the normalizing condition), but is singular at 
${\cal M} \to 0$ may have been subtracted from the final result.

\begin{figure}[thb]
\centerline{\psfig{figure=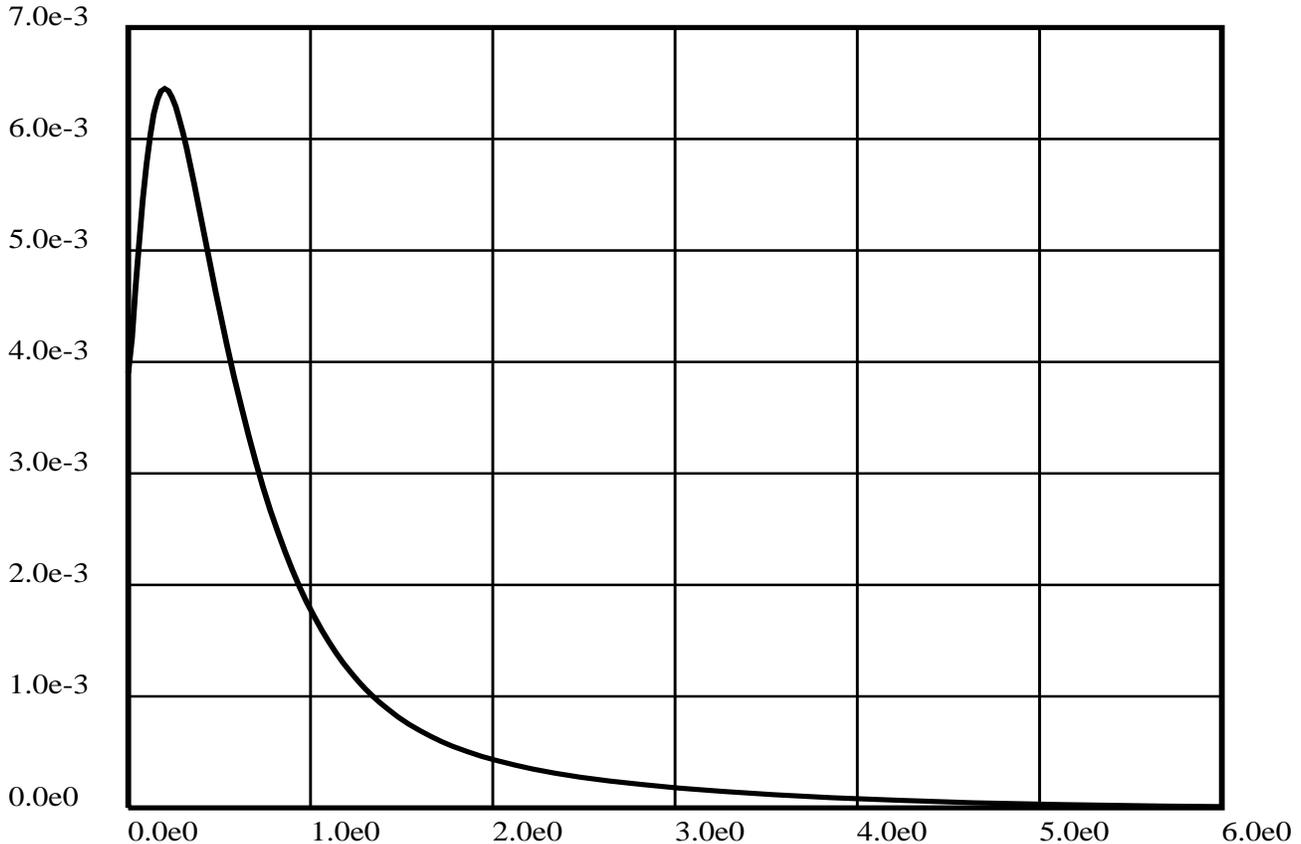,height=12cm,width=18cm}}
\caption{\label{sph_uni} The Casimir energy of a massive scalar field in 
the whole space with Dirichlet boundary condition on a sphere of radius 1 
as a function of the mass of the field.}
\end{figure}

As has been pointed out in \cite{bordag_1}, \cite{bordag_3} there is no 
argument at present which can remove this arbitrariness in case of a 
massless scalar field in the interior (or exterior) of a sphere. Therefore 
Casimir effect in the whole space with Dirichlet boundary conditions on 
the sphere is usually considered instead of interior alone (see the 
Introduction).  It seems instructive to calculate the Casimir energy in 
this specific case employing our technique.

In fact there are two ways to proceed to take exterior into account. The 
first one deals with the continuous spectrum and requires that all the 
regularized sums be replaced with the appropriate integrals containing the 
energy levels density in their integrands. The second way lets one work 
with discrete spectrum all the time.  To carry that out one should place 
the initial spherical shell into another sphere with the radius 
$R_{out}=kR_{in}$ where $k \ge 1$ and calculate the Casimir energy for 
the system bounded by the outer sphere with the boundary conditions on 
both of the spheres taken into account. For each finite $k$ the spectrum 
remains discrete and the developed technique can be applied without any
modification.  The required result can be achieved in the limit $k \to 
\infty$. In practice it turns out that for $k \geq k_0$, where $k_0$ is 
finite and depends on the required precision only, the result doesn't 
depend on $k$. It turns out that in the case under investigation the 
4-digit precision of the final result can be achieved with $k_0 \simeq 
10$.

The final result of the calculations is presented on Fig.~\ref{sph_uni}. 
Note that while the qualitative behavior of Casimir energy is the same as 
that obtained with methods employing explicit subtraction 
\cite{bordag_3}, there is no absolute coincidence. For example, for the 
massless field the result obtained with our method is ${\cal 
E}_{cas}({M=0}) = 0.0039$ while direct subtraction leads to ${\cal 
E}_{cas}({M=0}) = 0.0028$. Again, the remaining discrepancy can be related 
to the finite arbitrariness of the subtractive procedure.

\section{Conclusion}
To summarize, an efficient technique for numerical calculation of Casimir 
energy of quantized fields in the presence of logarithmical divergencies, 
owing to non-trivial boundary conditions, has been developed.  The 
advantages of the proposed method are its ideological simplicity and 
universality which let one apply it to a wide range of problems in which 
numerical values for energy levels can be obtained with the sufficient 
precision.  As has been demonstrated the results of its application to a 
number of problems appear to be quite reasonable, including the cases with  
curved boundaries.  As to possible disadvantages they turn out to be 
mostly technical: the required floating-point precision to carry out all 
the necessary calculations in realistic cases turns out to be quite high 
(from three to four times higher than the one realized in the standard C 
double type).

\section{Acknowledgments}
This work has been supported in part by the RF President Grant 1450.2003.2.
One of us (Il.M.) is indebted to the Organizers of QFTHEP'2004 for 
hospitality and financial support. We are grateful for the fruitful 
discussions to O.Pavlovski and I.Cherednikov.


\end{document}